\documentclass[12pt]{iopart}

\usepackage{graphicx}
\begin{document}

\title[]{Observation of triple-photon decay in positron-electron pair annihilation: a triple coincidence setup for the undergraduate laboratory}

\author{M.~E.~A.~Elbasher, W.~Erasmus, E.~A.~M.~Khaleel, J.~Ndayishimye, P.~Papka}
\address{Department of Physics, University of Stellenbosch, Private Bag X1, 7602 Matieland, South Africa.}
\ead{papka@sun.ac.za}

\begin{abstract}

The positron-electron pair annihilation in two photons is known for its numerous applications using PET scanners. The decay of Positronium (Ps) from a standard sealed source in more than two photons is less likely but can be observed with a relatively simple setup. The main goal of this experiment was to verify momentum and total angular momentum conservation principles at subatomic level through the Ps annihilation. The two spin configurations of Ps are produced with a $\beta$$^{+}$ source. The decay modes are identified using $\gamma$ ray spectroscopy techniques in a double and triple coincidence setup. Three germanium detectors (or two HPGe and one NaI(Tl) detectors) and a digital electronic system are required. 

\end{abstract}

\maketitle

\section{Introduction}

The back-to-back emission of two 511 keV photons originates from the annihilation of the spin anti-aligned configuration of the Positronium (Ps). It is often employed to demonstrate the law of momentum conservation in subatomic processes. In this paper we describe a relatively simple triple coincidence measurement to observe the spin-aligned configuration of Ps. 
A number of three High-Purity Germanium (HPGe) detectors, a digital electronics data acquisition system with at least 3 channels and a $\beta$$^{+}$ source are required.
This measurement offers advanced features for data analysis and is an interesting case to visualise effects of Compton Scattering.

\section{Physics and applications of Positronium}

A proton decays into a neutron within a $\beta$$^+$ unstable nucleus with the subsequent emission of a positron and a neutrino which conserve the charge and lepton numbers:

\begin{equation}\label{beta}
p \rightarrow n + e^+ + \nu.
\end{equation}

\noindent
Following a $\beta$$^+$ decay, the emitted positron slows down to eventually stop in the substrate and, at rest, binds to an electron, forming a Ps atom. The two particles are held together by the electromagnetic force with binding energy 6.8 eV. Ps was reported experimentally for the first time in 1951 by M. Deutsch~\cite{De51} and has some properties similar to the hydrogen atom which consists of one electron and one proton. The positron has identical mass, {\it m$_e$} = (511 keV), spin, {\it s} = $\frac{1}{2}$, but opposite charge of the electron. Two different hyperfine structure state configurations of Ps depend upon the spin alignment of the electron and positron. If the spin alignment is antiparallel, .i.e. $(e^{+}\uparrow e^{-}\downarrow)$, a singlet state results which is known as para-Ps ({\it p}-Ps) with total spin {\it S} = $\frac{1}{2}$ $-$ $\frac{1}{2}$ = 0. When the alignment is parallel, .i.e. $(e^{+}\uparrow e^{-}\uparrow)$, it results in a triplet state which is known as ortho-Ps ({\it o}-Ps) with total spin {\it S} = $\frac{1}{2}$ + $\frac{1}{2}$ = 1. In vacuum, Ps atoms are formed of approximately 75$\%$ {\it o}-Ps and 25$\%$ {\it p}-Ps~\cite{Mi09}.

Although energetically stable, the particle/anti-particle pair soon annihilates into photons, elementary particles of electromagnetic radiation with spin value {\it s} = 1. The resulting total angular momentum ({\it J}) in the final state equals the total spin of Ps due to the zero orbital angular momentum ({\it l} = 0) of both configurations. In the anti-aligned case, the total angular momentum in the final state must be {\it J} = 0, resulting in an even number 2{\it n} of photons to conserve the total angular momentum: {\it J} = {\it n}$\times$1 $-$ {\it n}$\times$1 = 0. The most favourable case is {\it n} = 1 where two photons are emitted back-to-back with identical energy to conserve the momentum in the centre-of-mass frame. In the spin-aligned case, the final total angular momentum must be {\it J} = 1 requiring an odd number 2{\it n} + 1 of photons: {\it J} = {\it n}$\times$1 $-$ {\it n}$\times$1 + 1 = 1. The three individual photons are characterised by a continuous energy spectrum with total energy equal to 1~022 keV and directions verifying the momentum conservation in the centre-of-mass frame. An elegant explanation of the conservation laws applying to the Ps annihilation is detailed in Ref.~\cite{Ha04} for further reading.

The probability of the annihilation channel quickly decreases with the increasing number of photons emitted; this is mostly the reason why {\it o}-Ps has a very long half-life, $142.03\pm0.02$ ns in vacuum \cite{Va02}. Thus, the occurrence of more than three-photon decay is unlikely and its observation requires extremely sensitive measurements.
Owing to its much shorter lifetime, 125~ps~\cite{Fe10}, {\it p}-Ps is weakly affected by its surrounding environment and decays mostly via the annihilation of the $e^{+}e^{-}$ partners. To the contrary, the relatively long lifetime of {\it o}-Ps allows for the positron to annihilate with a neighbouring electron into two photons in a process called pick-off annihilation. Note also that the conversion from {\it o}-Ps to {\it p}-Ps occurs through spin-orbit interaction between Ps and the colliding atom~\cite{Mi09,Fe10,Uj78}. It was also found that $\approx$ 2$\%$ of the positrons undergo in-flight annihilation while slowing down~\cite{He84}. As a result, the observed strength of the spin-aligned Ps formed in a gas or a solid is much shorter than the anti-aligned configuration.

Recent developments in the field of positron storage systems paved new avenues in the study of symmetries using Ps atoms or anti-hydrogen atoms~\cite{Im09}. The binding of Ps to molecules or to itself in Ps$_{2}$, the equivalent of H$_{2}$, was recently firmly observed~\cite{Ca07}. The annihilation of the electron-positron pair is used in a number of applications such as Positron Emission Tomography (PET). PET is a subatomic imaging technique which produces a three-dimensional image of functional processes in the body. A short-lived positron emitter is injected into a living organism in an appropriate form. With sufficient time, the positron source is absorbed and concentrated in the specific organism. The body afterwards is placed in the imaging scanner to map the tissue concentration. The PET scanner is a large solid-angle detector arrangement, finely segmented and designed to detect coincident pairs of photons from the positron annihilation. There are many applications of Positron Emission Tomography (PET) imaging in the industry, for example in diamond mining or the characterisation of dynamical flows, in medicine for colorectal, breast, and prostate cancers, coronary artery disease or Alzheimer and Parkinson brain conditions.

\section{Experimental Setup}

\begin{figure}[ht!]
 \centering
 \includegraphics[width=8cm]{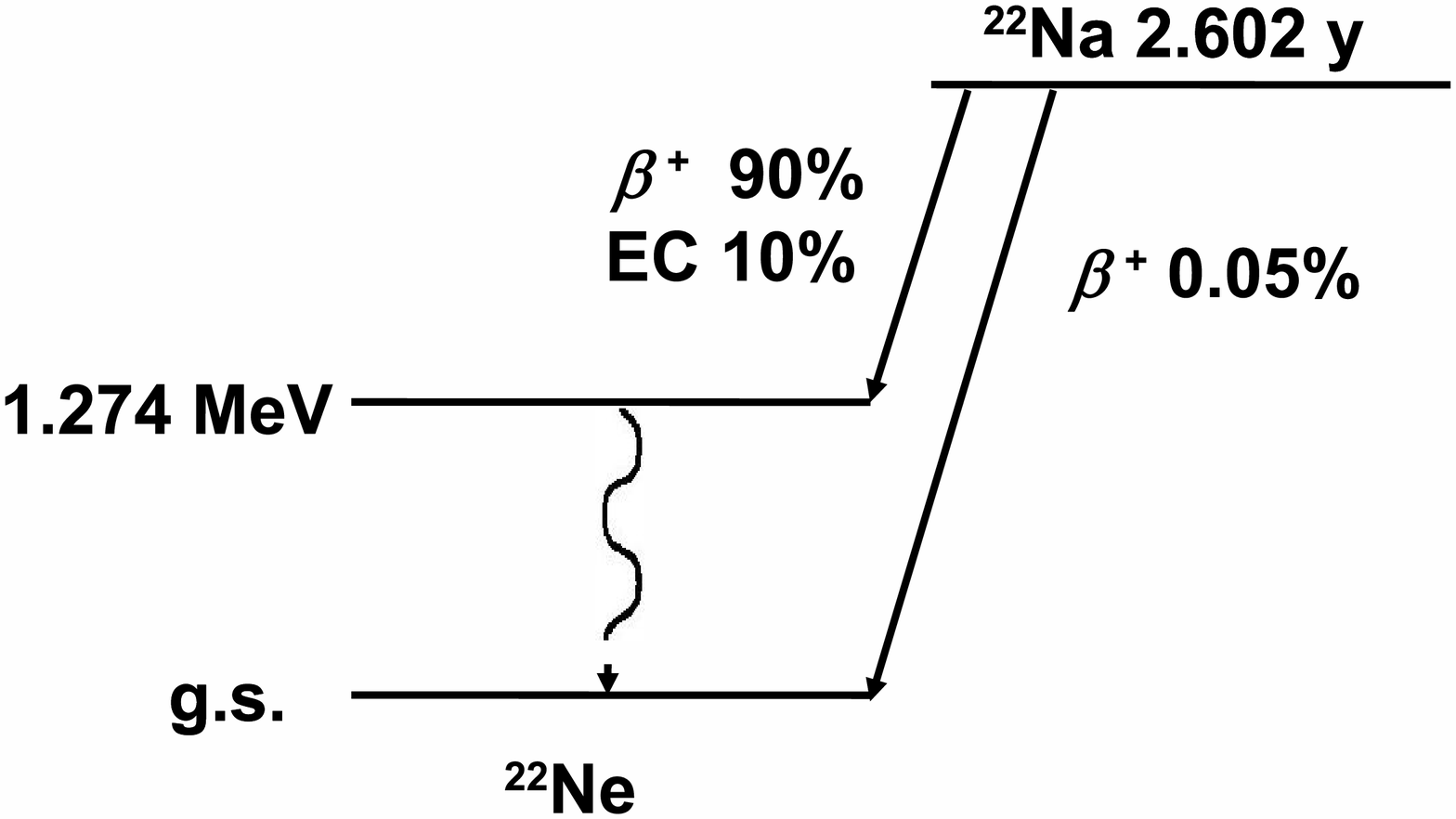}
 \caption{\label{na22} Decay Scheme of $^{22}$Na.}
 
\end{figure}

The aim of this experiment was to identify the two decay channels of the Ps annihilation. The back-to-back coincident measurement requires two detectors set in coincidence mode pointing at a $\beta$$^{+}$ source from two opposite directions. Difficulties arise in identifying the annihilation in three photons with a standard $\beta$$^{+}$ source owing to the rather low strength of this particular channel, compared to a relatively large amount of back-to-back 511 keV photons. In this experiment, a radioactive source of $^{22}$Na was used. As showed in Fig.~\ref{na22}, $^{22}$Na decays by positron emission or Electron Capture (EC) to the 1~274 keV excited state of $^{22}$Ne with a branching ratio of nearly 90$\%$ and 10$\%$ respectively and a rather low component of 0.05$\%$ decays directly to the ground state via $\beta$$^{+}$~\cite{Fi05}.

\begin{figure}[ht!]
 \centering
 \includegraphics[width=9.0cm,height=4.0cm]{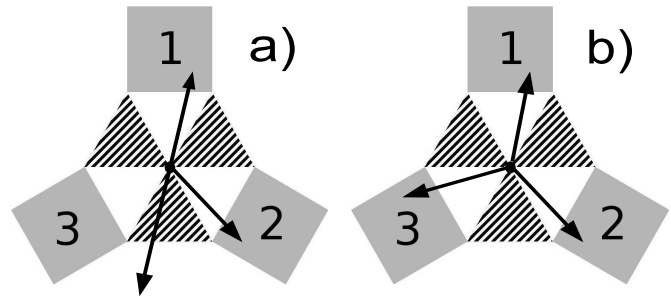}
 \caption{\label{set1} Three Germanium detectors at equally-spaced angles pointing at a $^{22}$Na source. a) Two back-to-back photons from {\it p}-Ps annihilation with one of the 511 keV photons and the 1~274 keV transition detected in two individual detectors. b) Triple-photon emission from {\it o}-Ps annihilation with the 1~274 keV $\gamma$ ray omitted.}
 
\end{figure}

\begin{figure}[ht!]
 \centering
 \includegraphics[width=9.0cm,height=4.0cm]{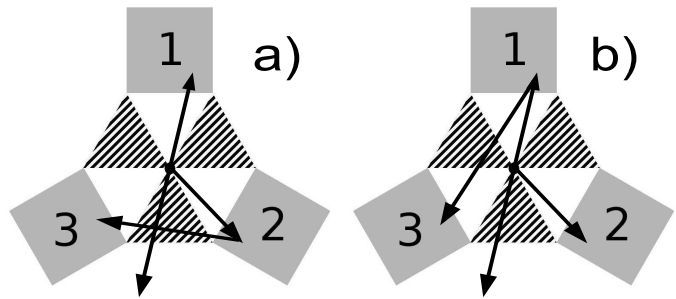}
 \caption{\label{set2} Three Germanium detectors at equally spaced angles pointing at a $^{22}$Na source. True three-fold events represented for two-step energy deposition of: a) the 1~274 keV transition and b) the 511 keV annihilation photon.
}
 
\end{figure}

\noindent
The experimental setup presented in Figs.~\ref{set1} and \ref{set2} is an arrangement of three Germanium detectors, noted 1, 2 and 3, each placed at the summit of an isosceles triangle pointing at a radioactive source located at the barycentre. The distance between the source and the detectors is chosen such that no straight line passes through the source and more than one detector. The logic lies purely on the concept of momentum conservation to suppress the {\it p}-Ps events drastically and preferentially select the triple-photon emission. Thus, if three photons are emitted in three different directions with approximately equal energy, the photons must be emitted on a plane with relatively similar angles between every direction due to momentum conservation. Since none of the detector pairs are aligned, it is impossible to detect the back-to-back photons in coincidence when the annihilation takes place in the vicinity of the source as shown in Fig.~\ref{set1}a). Only one 511 keV photon with the de-excitation $\gamma$ ray can be detected in coincidence. In Fig.~\ref{set1}b), three individual photons are favourably detected with a large solid angle represented by the empty triangles. Note that in Fig.~\ref{set1}b) the 1~274 keV transition is omitted; those events correspond to the $\beta$$^{+}$ decay towards the ground state or when the 1~274 keV transition does not hit a detector.

Two effects must be carefully considered. If the configuration in principle allows for exclusive detection of three photons, it must be remembered that the Compton Scattering occurs predominantly for those photons with {\it E}$_{\gamma}$ $>$ 500 keV. In this setup, the detectors are not shielded against Compton escape and a $\gamma$ ray is likely to deposit its energy in two individual detectors through Compton Scattering, as illustrated in Fig.~\ref{set2}. The energy of two photons can be deposited in three detectors, following some scattering combinations as in the two possible cases illustrated in Figs.~\ref{set2}a) and~\ref{set2}b). For those events, the total energy is expected to be 511~+~1~274~keV or less if one of the two photons Compton-scattered further and escaped. The equally-spaced arrangement of the detectors relies on a point-like source placed at the intersection of the three axis pointing at the centre of the detectors. However, if the encapsulation of the source is not sufficient, some positrons can travel a substantial distance in the air before being fully stopped. For those Ps annihilating between two detectors via back-to-back photons, a triple-coincidence event can be triggered through Compton Scattering, but in this case with a total energy equal to or less than 1~022 keV.

For this measurement it is advisable to use three HPGe detectors as the high resolution allows for very efficiently separating the photopeaks from the Compton background. The resulting total energy spectrum, being the energy sum from the three detectors, must be of sufficient resolution in order to resolve the peaks. Using three identical germanium detectors of resolution 0.17$\%$, the Full Width at Half Maximum (FWHM) of the energy sum is equal to: 

\begin{equation}\label{reso1}
FWHM_{total}=\sqrt{{FWHM_{A}^{2}}+{FWHM_{B}^{2}}+{FWHM_{C}^{2}}} \approx  0.3 \%.
\end{equation}

The triple coincidence can naturally be set using standard NIM modules. This option can be very educative but it requires a relatively sophisticated data acquisition system if the energy and time of the three detectors are to be recorded. We chose to perform this experiment with a 4-channel PIXIE-4 digital electronic system from XIA~\cite{xia0}. The trigger is set via software with adjustment of the coincidence level and time window. Time stamp and energy information are written event by event in a binary file. In this experiment, three 2" planar germanium detectors were placed at a distance of 8~cm from a relatively weak source of $^{22}$Na ($\approx$~0.5 $\mu$Ci), resulting in a count rate of about 1.5 triple-photon decay events per hour (150 events in four days). The counting time can be reduced to less than 24 hours with a source of about 5 $\mu$Ci, but should not be stronger for count rate limitation and increased accidental coincidence rate reasons.

\section{Data Analysis and Results}

\begin{figure}[ht]
 \centering
 \includegraphics[width=13cm]{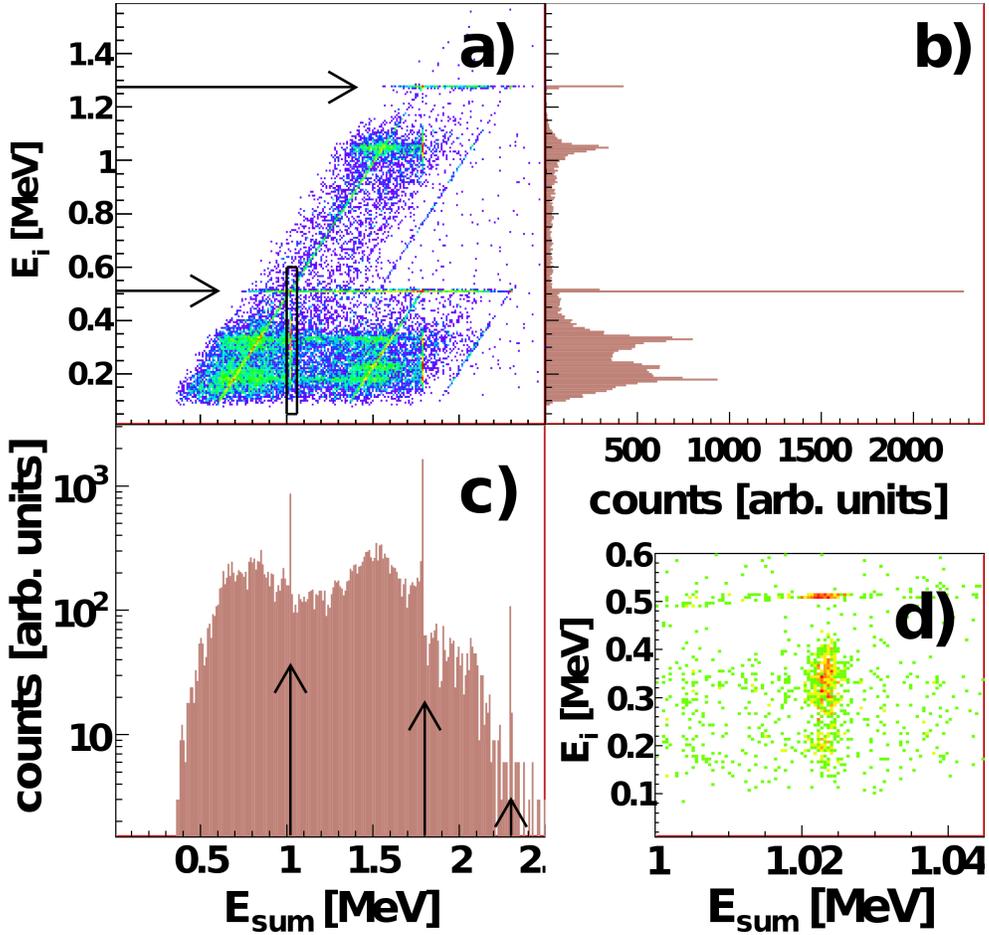}
 \caption{Panel a): 2D scatterplot of the energy measured in one detector ({\it E}$_{i}$, Y-axis) versus the total energy measured in the three detectors ({\it E}$_{sum}$ = {\it E}$_{1}$ + {\it E}$_{2}$ + {\it E}$_{3}$, X-axis). The solid rectangle indicates the region enclosing the locus of the triple-photon events. Panel b): 1D-projection of {\it E}$_{i}$; Panel c): 1D-projection of {\it E}$_{sum}$; Panel d): zoom on the three photon locus. }
\label{2d} 
\end{figure}

The three cases of triple coincidence detailed earlier are clearly observed in the 2D scatterplot of Fig.~\ref{2d}a). The energy {\it E}$_{i}$ of one detector ({\it i} = 1, 2 or 3) on the Y-axis is plotted against the total energy {\it E}$_{sum}$ = {\it E}$_{1}$ + {\it E}$_{2}$ + {\it E}$_{3}$ recorded in the three detectors. The 1D projections of {\it E}$_{i}$ and {\it E}$_{sum}$ are represented in Fig.~\ref{2d}b) and Fig.~\ref{2d}c) respectively. The two peaks indicated by the arrows at {\it E}$_{sum}$ = 1~022 and 1~785 keV in Fig.~\ref{2d}c), represent the events where the total energy of two photons is fully deposited in three detectors. The {\it E}$_{sum}$ = 1~785 keV peak is the energy sum of the 1~274 keV plus a single 511 keV photon corresponding to the scenarios of Fig.~\ref{set2}. The {\it E}$_{sum}$ = 1~022 keV peak is a combination of the two annihilation channels of {\it o}-Ps and {\it p}-Ps. The higher energy peak at {\it E}$_{sum}$ = 2~296 keV is the energy sum of the two back-to-back 511 keV photons plus the 1~274 keV transition corresponding to the decay of the Ps away from the source between a pair of detectors with total energy absorption in three detectors.

The horizontal and oblique lines in the 2D scatterplot of Fig.~\ref{2d}a) correspond to incomplete events where at least one photon has  fully deposited its energy while the other photons scattered and only part of the energy was measured. The horizontal lines indicated by the solid arrows in Fig.~\ref{2d}a) are the photopeak energies of the 511 and 1~274 keV photons in one detector; part of the energy of the other photon(s) is lost.

\begin{figure}[ht]
 \centering
 \includegraphics[width=5cm,height=5cm]{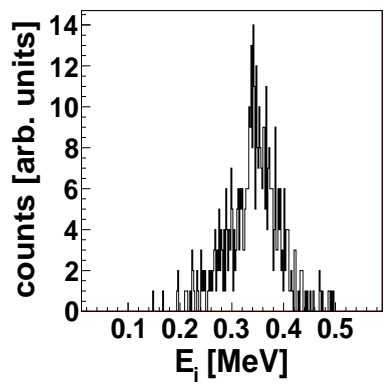}
 \caption {Energy spectrum of a single photon measured in the triple-photon annihilation of Ps.}
\label{Eg} 
\end{figure}

The candidates for the triple-photon emission are selected from the matrix of Fig.~\ref{2d}a) by placing a gate on the total energy peak {\it E}$_{sum}$ = 1~022 $\pm$ 2 keV, with the condition that no detector, or combination of two detectors, measures a photopeak of 511 keV. The energy of a single photon emitted from the triple-photon annihilation after selection is shown in Fig.~\ref{Eg}. The mean energy of the photon is approximately a third of the available energy {\it E}$_{average}$ $\approx$ 340 keV. This is consistent with the geometry of the experimental setup selecting events where the photons are emitted with a similar angle $\Delta$$\theta$~$\approx$~120$^{\circ}$ with respect to each other. Note also that the three-body kinematics and finite opening angle of the detectors cause a continuum energy distribution of the single photons.

The likelihood of two 511 keV photons emitted away from the source mimicking the {\it o}-Ps events through Compton scattering of both photons is questionable. In this scenario, one detector is hit twice in a sufficiently short time for the double hit to be recorded as one and the amplitude of the signal to correspond to the energy sum of the two photons. Consequently, the signature of the 511 keV peak is lost and cannot be used to discriminate the {\it p}-Ps events. A measurement was performed with lead shields placed between the detectors to eliminate such cross talk events. As a result, the 511 keV photons were nearly fully suppressed but the triple-photon emission was still observed. 
A measurement was also performed with two HPGe detectors and one NaI(Tl) detector to investigate the effect of a poorer resolution. The total resolution was nearly equal to the resolution of the single NaI(Tl) but the triple-photon decay could still be isolated. In such a case it is advisable to make use of some shielding as the rejection of unwanted events becomes less effective with poor resolution.

\section{Conclusion}

In this work the triple-photon coincidence measurement using a $^{22}$Na was performed. A number of important aspects in experimental nuclear physics were encountered, for example the Compton scattering visualised through the detection of a single transition in two individual detectors. A powerful way of identifying total energy events was illustrated using 2D scatterplots. Such a technique is widely used in nuclear physics to isolate events in exclusive coincident measurements. The observation of Ps annihilation was observed from its two configurations, {\it o}- and {\it p}-Ps by means of two- and three-photon coincident measurement. This verified not only the momentum conservation but also the total angular momentum conservation in {\it o}-Ps and {\it p}-Ps annihilation.

\bigskip
\noindent
{\bf Acknowledgments}
\bigskip

We thank H. Schwoerer for the thoughtful discussions and for triggering the idea for this experiment. It is also a pleasure to thank S. M. Wyngaardt and J. J. van Zyl for reading the draft of this paper and providing comments. This work was financially supported by the National Research Foundation of South Africa.

\end{document}